# Multiagent-based Participatory Urban Simulation through Inverse Reinforcement Learning

**Soma Suzuki**

Center for Advanced Spatial Analysis
University College London

## Abstract

The multiagent-based participatory simulation features prominently in urban planning as the acquired model is considered as the hybrid system of the domain and the local knowledge. However, the key problem of generating realistic agents for particular social phenomena invariably remains. The existing models have attempted to dictate the factors involving human behavior, which appeared to be intractable. In this paper, Inverse Reinforcement Learning (IRL) is introduced to address this problem. IRL is developed for computational modeling of human behavior and has achieved great successes in robotics, psychology and machine learning. The possibilities presented by this new style of modeling are drawn out as conclusions, and the relative challenges with this modeling are highlighted.

## I. Introduction

Urban modeling is traditionally seen as rational and technical (Sager 1999), thus leading to massive failures in urban planning (Jacobs 1961). To interpret knowledge of local context (McCall 2003), participatory simulation has been studied intensively, by allowing experts and non-experts to interactively define the model (Drogoul, Vanbergue and Meurisse 2002). In conjunction with participatory simulation, multi-agent human movement modeling has garnered attention in that the model enables bottom-up simulation which has the potential to deemphasize the manipulative intrusion (Epstein 1999). Even though it made great progress, the limitation of the multi-agent simulation in constructing highly realistic agents has been the subject of much discussion (Batty and Torrens 2001). The incurred distrust in the simulation model has always vexed urban planners and impeded efficient urban planning. The constraints mainly arise from the number of variables that the agent needs to address since the solution space quickly becomes intractable as the number of parameters increases. A promising approach for overcoming this drawback is to learn preferences of humans from the observed demonstration by inverting a model of rational decision making given a reward function (Russell and Norvig 1995). This approach is known as Inverse Reinforcement Learning (IRL) in Markov Decision Process (MDP) (Ng and Russell 2000) in the machine learning community.

This paper attempts to address the limitation of multi-agent participatory simulation when replicating social phenomena. In section II, the importance of participatory simulation is discussed with a particular experiment in the city of Kyoto. To demonstrate that IRL creates another paradigm in urban simulation, the contribution and the limitation of existing theoretical models are examined in section III. Then Markov Decision Process and Inverse Reinforcement Learning are introduced in section IV with a specific example of robot simulation to illustrate the potential of IRL. Finally, the conclusion and future work are discussed in section V.

## II. Participatory Simulation

Significant planning difficulties result from the estrangement between the stakeholder's local knowledge and planner's technical theory. With the participation of the stakeholder in the process of urban planning, more plausible solutions are expected to be achieved. This is particularly true of multi-agent urban simulation design. It is most unlikely that a virtual agent merely based on academic data reflects local knowledge.

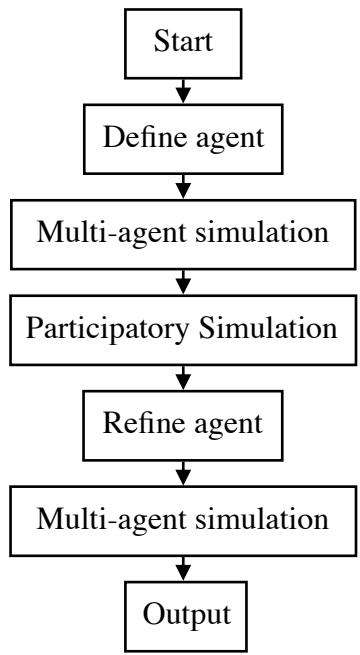

Figure 1. Participatory design.

The recent rise of participatory urban simulation has opened a new field of research to address this problem, by enabling the model to incorporate knowledge of both experts and non-experts, through a role-playing game or by being immersed in the simulation as "human agent" (Edmonds 2000). The human agents interact with virtual agents to gain the insight into the simulated model. The brief process of participatory urban modeling is illustrated in Figure 1 (Ishida, Nakajima, Murakami and Nakanishi 2007). The learning process of multiagent-based participatory simulation consists of two phases: the deductive agent design and participatory agent design. At first, experts generate agents based on the domain knowledge and data. After validating the model, the human agents are immersed in the simulated environment and the participatory simulation is performed. There is also an extension of participatory simulation known as argument experiment, where an experiment is performed in real space by humans with a multiagent virtual simulation. Take, for instance, the augmented experiment has been conducted in the city of Kyoto (Ishida, Nakajima, Murakami and Nakanishi 2007).

*Example: Augmented Experiment in Kyoto*

To simulate disaster evacuation in Kyoto station, a tracking system that navigates passengers based on their current location was deployed. Beyond the conventional evacuation system in stations, which announces route information from public megaphones, the systems are intended to instruct individuals via mobile phones. In the augmented experiment, the pedestrians' movements were captured using GPS trace. Then, the real-time human movements were projected onto avatars in virtual space. In parallel with the real space experiment, a multiagent-based simulation with a large number of agents generated by domain theory was conducted. Finally, the system instructed the evacuees to evacuation destinations, with a) two-dimensional map and b) three-dimensional virtual spaces. Ten humans and three thousand agents undertook the experiment. Through the experiment, it turns out that a map is an excellent method for evacuee navigation. On the other hand, the small and low-resolution images of three-dimensional virtual spaces on mobiles phones were not very effective.

By integrating the real world experiment and virtual world experiment, the experiment is augmented, enabling large-scale simulation with the small number of subjects for experiments. By incorporating theory based agents and human-controlled agents, multiagent-based participatory urban simulation is expected to fill in the gap between the refined academic theory and practical domain knowledge, and an expected consequence being that stakeholders are more engaged in the planning process to achieve more effective urban modeling.

## III. Limitation on Agent Design

Models are, by definition, representation of a simplification of the real world, and consequently, any simulation models are necessarily incomplete and partial (Batty and Torrens 2001). However, they need to, of course, achieve minimum plausibility to be utilized in actual planning. There are two issues that we should draw together to illustrate the limitation of reproducing the real world with traditional agent-based pedestrian simulation. The first is the issue of the complications that arise from the rule-based models which induce intractable mathematical operations. This leads to the second issue of the impracticability of validating all the plausible models against data. The archetypal example of this is an agent-based pedestrian model *STREETS* that Center for Advanced Spatial Analysis in University College London developed (Schelhorn, O'Sullivan, Haklay and Thurstain 1999).

*Example: STREETS*

In the simulated environment of *STREETS*, the street network with various attractions such as shops, offices and public buildings are configured. Each pedestrian agent was attributed two characters: socio-economic and behavioral. Socio-economic characteristics contribute to income and gender, which define the locations that the agents are intended to visit. Behavioral characteristics dictate the detailed behavior of agents such as speed, visual range, and fixation. The speed is the maximum walking speed of the agent. Visual range defines which elements in the environment the agent will see. Fixation relates to how consistent the agent is in the pre-determined activity schedule upon encountering the configured attraction. Finally, various parameters are introduced to determine the likelihood of following distractions. a) The match of agent types and building types. For instance, an agent who is engaged in the activity of shopping is assumed to enter into another shop during the journey but is less likely to be attracted by the office building. b) The attractiveness of the building for a certain demographics. c) The level of inconsistency of the agent, which is a general behavioral characteristic defined as fixation. This models the reality that some people are more easily distracted than others. Fixation interacts with the agent's internal clock, which measures the time an agent is supposed to stay in the system. As the remaining time decreases, the agent becomes more concentrated in completing the pre-determined schedule.

Even though the STREETS is an interesting approach toward modeling social phenomena in the city and explains the pedestrian movement to a certain level, the generated simulation is not realistic enough to be used in policy-making. First and most importantly, pedestrians' activity schedule and the possible deviation when finding the attractive building is not, of course, based merely on their gender and salary. A naive solution to replicating the real world more plausibly is to use more variables, or rules such as age, ethnicity and household structures, with more detailed parameters on the attractiveness of facilities. However, the solution space explodes enormously as the number of variables increases, and there are almost infinite numbers of factors involved in a pedestrian activity and incorporating all the elements is typically intractable. To make matters worse, even supposing all the necessary variables are described, there is no way that all the elements of the model are validated. This dilemma between the parsimony model and complex system has been the chief limitation of the agent-based pedestrian model in practical applications, and a consequence of this is that the simulation model is relegated to non-policy contexts such as education, a group discussion, learning from visualization and even entertainment (Batty and Torrens 2001). Although those applications of the model are very important, the original passion of replicating the reality is seemingly lost.

## IV. Inverse Reinforcement Learning

Instead of merely reacting to the environment, the cognitive agents need to take actions based on what they individually observe. In this section, Inverse Reinforcement Learning (IRL) and Markov Decision Process (MDP) are briefly introduced as the potential use for generating realistic and plausible pedestrian agents.

*Markov Decision Process*

A finite MDP is a tuple ($S$, $A$, $P_{sa}$, $\gamma$, $R$), where

- $S$ is a finite set of $N$ states $\{s_1, s_2, …, s_n\}$.
- $A$ is a set of $K$ actions $\{a_1, a_2, …, a_k\}$.
- $P_{sa}$ is state transition probabilities upon taking action $a$ in state $s$.
- $\gamma \in [0,1]$ is the discount factor.
- $R : \mathbb{S} \times \mathbb{A} \to \mathbb{R}$ is the reward function that depends on state and action.

The classical problem of MDP consists of finding the optimal policy $\pi^* : \mathbb{S} \to \mathbb{A}$ which selects the actions that maximize the expected reward for every state.

*Inverse Reinforcement Learning*

Inverse Reinforcement Learning (IRL) problem is to find the reward function from an observed policy. More specifically, given a finite space $S = \{s_1, s_2, …, s_n\}$, set of actions $A = \{a_1, a_2, …, a_k\}$, transition probabilities $P_{sa}$, a discount factor $\gamma \in [0,1]$ and a policy $\pi : \mathbb{S} \to \mathbb{A}$, the goal of IRL is to find a reward function such that $\pi$ is an optimal policy $\pi^*$. When learning reward functions through IRL, the ultimate objective is to guide an artificial agent with the acquired reward function, and a consequence being that the agent learns a "good" policy or simply imitates the observed behavior. IRL is a very promising approach in generating pedestrian agents in the urban environment as the current pervasiveness of mobile devices allows us for the collection of mobile location data such as GPS, Wi-Fi, or RFID, namely training data at an unprecedented scale and granularity (Mehrotra and Musolesi 2017).

The traditional agent-based pedestrian model assumes, in a sense, that the reward function is known: *STREETS* assumes that the attractiveness, in other words, the reward of each building, is fixed. However, it seems evident that in examining human behavior the reward function must be considered unknown to be ascertained through inductive reasoning (Ng and Russell 2000). As it is infeasible to determine the relative weights of the factors such as the desire for shortest path, avoidance of crowded areas and movement with the flow, IRL is a fundamental problem of the agent-based pedestrian model. To give a clear example, the successful application of IRL in robot navigation is illustrated here.

*Example: Robot Navigation with IRL*

In the context of socially intelligent robot navigation, a human-like walking pattern has been investigated (Shiarlis, Messias and Whiteson 2017). The experiments took place in a simplified social environment where two people are placed randomly in a room. The purpose of the experiment is to find the reward function that accounts for the relationship between the placed people and the navigation route. For example, if people are facing each other, the robot should not pass between them as they are likely engaged in some activities such as conversation. In contrast, if they are looking away from each other, it might be better to navigate between them if it yields the shortest path. With the trajectory data of human demonstrator for multiple initial and final points, the reward function is learned and then deployed in the robot. As a result, the robot demonstrated a plausible and socially compliant path.

Learning socially intelligent navigation and generating realistic pedestrians in urban environments is intuitively equivalent in that both processes address the problem of how humans "would" behave. In light of the success in robotics, it is not altogether irrelevant to attempt anew to replicate pedestrians in the city. Although IRL is a promising approach for learning pedestrian movement patterns, there are several challenges that need to be addressed. Furthermore, as the model does not need a particular underlying theory, it could be applied to scenarios where no theory has yet been established and even possibly used to develop new human behavior theory.

*Pitfall in Inverse Reinforcement Learning*

There are three conceivable issues for generating pedestrian agents through IRL. As a full description of the pitfalls of IRL is beyond the scope of this paper, only the major challenges are introduced briefly. a) As is always the case with deductive models, a quantitative evaluation is difficult since no ground truth data is available. Instead, the model is qualitatively validated. The plausible validation method needs to be considered. b) The other problem arises as the learned reward functions do not necessarily explain human preferences. People often deviate from optimality (Evans, Stuhlmuller, and Goodman 2015). To illustrate, if an agent repeatedly fails to choose preferred actions due to cognitive bias or irrationality, the model concludes that the agent does not prefer the option at all. To illustrate, a number of people smoke every day while intending to quit and considering the action to be regrettable. c) In addition, a major challenge in applying reinforcement learning in the multiagent model is how to manage the explosive computational cost as the state-action space grows exponentially with the number of agents and the learning becomes prohibitively slow. There are two possible approaches to alleviate this drawback. Firstly, as known as structural reinforcement learning, it is possible to reduce the size of the state-action space by supplying the model with the partial, but fundamental pedestrian movements in the form of connectivity graph structures (Hillier 1989). Secondly, by replacing the simulation model with a fast statistical surrogate (also called an emulator), the state-action space is more efficiently explored, thus leading to less expensive computation. To achieve this, one needs to apply a dimension reduction technique to the input space, as done for Gaussian Process emulators for instance (Liu and Guillas).

## V. Conclusion

In this paper, Inverse Reinforcement Learning (IRL) is presented for designing multiagent-based participatory urban simulation. This has helped to demonstrate that the existing agent-based urban modelings, although they are declared as "agent-based", are invariably top-down and tautological, being less informative of real world, This is due, for the most part, to the intractability of factors involving human behavior, and to the wrong attempt to dictate them. An interdisciplinary approach to spatial analysis, robotics, and computational psychology is required to realize robust simulation for efficient urban planning. Through the immediate feedback from the model that incorporates knowledge of experts and non-experts, urban planning is expected to be more nimble, responsive and effective. More importantly, as qualitative validation is not available in participatory urban simulation, the acquired simulation result needs to be qualitatively interpreted. Consider, for example, that the pedestrian simulation in the urban environment is performed and a number of accidents have occurred in the virtual environment. With the lack of fear of death, participated stakeholders would behave in a fundamentally different way. To gain plausible insights into social phenomena from participatory simulation, a standardized validation method needs to be established: subsequent examination on the levels of reality participants experienced is one of the possible analysis.